# The optical variability of QSOs *

S. Cristiani[1], S. Trentini[1], F. La Franca[1], I. Aretxaga[2], P. Andreani[1], R. Vio[1,3], and A. Gemmo[1]

[1] Dipartimento di Astronomia dell'Università di Padova, Vicolo dell'Osservatorio 5,
I-35122 Padova, Italy
[2] Royal Greenwich Observatory, Madingley Road, Cambridge CB3 0EZ, U.K.
[3] ESA, IUE Observatory, Villafranca del Castillo, Apartado 50727, 28080 Madrid, Spain

**Abstract.** The long-term variability of a sample of 180 optically selected QSOs in the field of the Selected Area 94 has been studied. The relations between variability and luminosity and between variability and redshift have been investigated by means of statistical estimators that are "robust" and allow at the same time to eliminate the influence of the measurement errors. A comparison is carried out with the results of two other samples of QSOs, in the fields of the South Galactic Pole (Hook et al. 1994) and of the Selected Area 57 (Trevese et al. 1994). Merging the three samples provides a total of 486 QSOs. The analysis in the QSOs rest frame of both the ensemble structure function ($SF$) and the individual variability indices show that: 1) a negative correlation between variability and luminosity is clearly present, in the sense that more luminous QSOs show less variability; 2) a significant positive correlation exists between variability and redshift; 3) such correlations may be equally well parameterized either with a model in which the timescale of the variability is fixed for all the QSOs ($\tau \sim 2.4$ yr), while the amplitude linearly increases with the absolute magnitude and redshift, or with a model in which the timescale of the variability linearly depends on the absolute magnitude and the amplitude is only a function of the redshift.

The same analysis carried out in the observer's frame provides the following results: 1) there is a negative correlation between variability and luminosity; 2) the timescale of variability does not depend significantly either on the absolute magnitude or on the redshift; 3) the ensemble structure function is well represented by a parameterization in which, with a fixed timescale of about 5.5 yr, the amplitude linearly increases with the absolute magnitude; 4) although the general behaviour of the $SF$ does not show a systematic variation of the timescale and/or amplitude with redshift, if we examine the average variability index for objects with $-25 > M_B > -27$, we find that below redshift 1 quasars are significantly less variable than at higher redshift.

The implications in terms of the black-hole, starburst and microlensing models are briefly discussed.

## 1. Introduction

Variability is a common property of QSOs. Although not thoroughly understood and exploited, it provides a potentially powerful tool to constrain theoretical models. A number of different scenarios, such as the black-hole (Rees 1984), the starburst (Terlevich et al. 1992), the gravitational microlensing (Hawkins 1993) models, in principle can be tested on the basis of their predictions about the temporal variations of the QSO luminosity. Attempts in this sense have been reviewed by Wallinder et al. (1992) in the framework of the black-hole model. Up to now, unfortunately, the results have not come up to general expectations. Even the determination of simple phenomenological trends, such as the correlations of variability with redshift or luminosity, has resulted in a complex situation (Cristiani et al. 1990, Giallongo et al. 1991, Hook et al. 1994, Trevese et al. 1994, Paltani & Courvoisier 1994).

Other important results from the study of the ensemble variability of QSOs concern the understanding of the biases affecting samples selected on the basis of the sole variability (Trevese et al. 1989, Véron & Hawkins 1995) or of multicolor data (Warren et al. 1991). The analysis of the relationship among the luminosity in different bands of the QSO e.m. spectrum is also affected by variability (e.g. the $L_x - L_o$ relationship, La Franca et al. 1995).

Variability provides an easy way to select independent QSO samples to cross-check the biases of various selection techniques (Cimatti et al. 1993, Trevese et al. 1994).

standing at least of the above mentioned phenomenological trends prompted us to continue and refine the study of the field of the Selected Area 94 (Cristiani et al. 1990), developing statistical techniques allowing a meaningful comparison with the other major variability studies (Hook et al. 1994, Trevese et al. 1994).

In the following we report in Section 2 the description of the QSO of the Selected Area 94 (SA 94) sample and the photometric material used to study its variability; in Section 3 the characteristics of two other samples in the Selected Area 57 (SA 57) and South Galactic Pole (SGP) are recalled; in Section 4 the ensemble structure function is introduced as a statistical measure of the variability properties and the detected trends are further analysed with the aid of another variability index; the results and implications are discussed in Section 5.

Throughout the paper we have assumed cosmological constants $H_o = 50$ Km sec$^{-1}$ Mpc$^{-1}$, $q_o = 0.5$.

## 2. The SA94 sample

### 2.1. Definition of the sample

The SA94 sample is made up of 180 optically selected QSOs [1] (listed in Table 1), observed with 21 plates covering a time-base of 10 years. The QSOs belong to the UVx and objective-prism surveys carried out by Cristiani et al. (1991), La Franca et al. (1992), Cristiani et al. (1995); who also discussed the properties of the surveys in terms of selection criteria, completeness (better than 90%) and detection efficiency. 47 previously unpublished QSOs have been included in Table 1 in order to enlarge the sample and increase the statistical significancy. They have been selected with the same criteria from the same databases as the rest of the sample and therefore are expected to share the same variability properties. Their identification has been carried out with spectroscopic observations obtained at the ESO 3.6-m telescope in La Silla, equipped with the OPTOPUS multifibre spectrograph (Avila & D'Odorico 1993), in three observing runs on September 1990 and September and October 1991. The OPTOPUS multifiber spectrograph uses bundles of 50 optical fibers, which can be set within the field of the Cassegrain focal plane of the telescope; this field has a diameter of 32 arcmin, and each fibre has a projected size on the sky of 2.5 arcsec. We used the ESO grating #13, providing a dispersion of 450 Å mm$^{-1}$ in the wavelength range from 3800 to 9800 Å. The resolution was 27 Å. We dedicated about half of the fibers

---

[1] In the present paper QSOs are defined as objects with a starlike nucleus, broad emission lines, brighter than $M_B = -23$ mag, applying the K-corrections computed on the basis of the composite spectrum of Cristiani and Vio (1990), and dereddened for galactic extinction according to Burstein and Heiles (1982).

**Table 1.** List of the QSOs in the SA94

| $\alpha$ | $\delta$ | z | $B'$ | $IDX$ | Ref. |
|---|---|---|---|---|---|
| 2 43 30.42 | −0 44 46.2 | 1.596 | 18.80 | +0.07 | a |
| 2 43 33.27 | −1 08 06.2 | 1.422 | 20.43 | +0.03 | e |
| 2 43 36.30 | −1 52 08.1 | 1.845 | 19.86 | +0.04 | b |
| 2 43 36.85 | +1 10 50.3 | 1.591 | 18.76 | +0.24 | e |
| 2 43 43.49 | −0 18 36.4 | 1.139 | 18.71 | +0.13 | e |
| 2 43 47.46 | −0 14 25.5 | 1.292 | 18.95 | +0.06 | e |
| 2 43 54.36 | +1 27 42.9 | 1.904 | 19.46 | +0.16 | e |
| 2 43 55.42 | −0 57 30.5 | 2.103 | 19.97 | +0.19 | e |
| 2 43 58.80 | −0 07 03.9 | 1.305 | 19.33 | +0.16 | a |
| 2 43 59.21 | −0 44 47.3 | 2.147 | 19.06 | +0.04 | a |
| 2 43 59.35 | −0 47 12.2 | 1.726 | 19.82 | +0.09 | e |
| 2 44 01.14 | +1 16 08.9 | 2.032 | 19.37 | +0.00 | a |
| 2 44 02.07 | −0 21 23.3 | 1.815 | 18.61 | +0.08 | a |
| 2 44 04.84 | −0 53 38.2 | 0.859 | 20.43 | +0.01 | e |
| 2 44 06.25 | −0 15 09.7 | 2.315 | 19.97 | +0.21 | e |
| 2 44 06.70 | −1 40 04.3 | 1.921 | 20.34 | −0.05 | e |
| 2 44 10.58 | −1 44 22.1 | 0.506 | 19.16 | +0.03 | e |
| 2 44 12.19 | −2 02 22.3 | 1.552 | 20.50 | +0.07 | e |
| 2 44 14.86 | −1 58 07.0 | 1.784 | 18.33 | +0.00 | a |
| 2 44 17.87 | −0 57 29.3 | 2.172 | 19.64 | −0.03 | e |
| 2 44 19.04 | −1 12 03.0 | 0.467 | 17.05 | −0.01 | c |
| 2 44 22.41 | +1 46 40.2 | 1.945 | 19.29 | +0.26 | a |
| 2 44 27.02 | −2 03 48.0 | 0.920 | 19.16 | +0.04 | e |
| 2 44 27.45 | −0 09 01.2 | 2.137 | 20.07 | +0.00 | e |
| 2 44 55.50 | −1 42 39.3 | 0.741 | 20.20 | +0.11 | e |
| 2 45 06.14 | −1 04 50.9 | 2.125 | 20.06 | +0.07 | e |
| 2 45 14.58 | −1 00 39.2 | 1.918 | 19.78 | +0.09 | e |
| 2 45 15.22 | −0 14 16.8 | 1.859 | 20.13 | +0.02 | e |
| 2 45 16.43 | +1 19 20.4 | 2.310 | 19.64 | +0.11 | c |
| 2 45 18.18 | −1 52 47.8 | 1.474 | 20.32 | +0.23 | e |
| 2 45 20.98 | −1 44 54.0 | 1.937 | 19.03 | +0.03 | a |
| 2 45 22.87 | −0 28 24.9 | 2.118 | 18.52 | −0.02 | a |
| 2 45 27.96 | −0 52 44.3 | 0.812 | 19.71 | +0.26 | e |
| 2 45 33.58 | +0 23 23.9 | 0.835 | 19.82 | +0.00 | e |
| 2 45 41.80 | +0 16 49.0 | 1.030 | 20.50 | −0.01 | e |
| 2 45 47.58 | −0 38 14.4 | 1.450 | 19.86 | +0.05 | e |
| 2 45 49.38 | +0 23 25.5 | 1.015 | 20.07 | +0.04 | e |
| 2 45 57.55 | +0 37 06.0 | 1.598 | 19.78 | +0.36 | e |
| 2 46 00.83 | −0 58 43.3 | 1.822 | 18.76 | −0.01 | a |
| 2 46 05.12 | +2 11 16.9 | 1.267 | 18.67 | +0.00 | e |
| 2 46 07.49 | −0 24 55.4 | 1.684 | 19.01 | +0.07 | e |
| 2 46 07.59 | −0 48 15.0 | 2.239 | 19.13 | +0.05 | e |
| 2 46 13.67 | −0 57 29.4 | 1.704 | 20.35 | +0.04 | e |
| 2 46 21.63 | +0 10 40.8 | 1.017 | 20.11 | +0.32 | e |
| 2 46 23.61 | −1 08 30.1 | 1.709 | 19.18 | +0.00 | e |
| 2 46 33.53 | −1 46 34.1 | 1.152 | 19.34 | +0.50 | e |
| 2 46 33.65 | −0 19 14.8 | 2.249 | 19.77 | +0.06 | e |
| 2 46 47.09 | +1 56 38.4 | 1.953 | 19.40 | +0.15 | a |
| 2 46 50.66 | −1 55 39.2 | 1.434 | 19.47 | +0.06 | e |
| 2 46 52.63 | −0 32 13.1 | 2.475 | 19.89 | +0.28 | e |
| 2 46 54.28 | +0 57 00.4 | 0.954 | 18.83 | +0.14 | a |
| 2 46 55.79 | −0 33 28.5 | 1.419 | 19.12 | +0.11 | e |
| 2 46 59.51 | −1 47 01.9 | 2.337 | 19.90 | +0.13 | e |
| 2 47 09.68 | +1 41 16.8 | 2.690 | 19.21 | +0.20 | a |
| 2 47 09.99 | +0 20 54.1 | 1.480 | 20.13 | +0.06 | e |
| 2 47 10.52 | +1 10 27.3 | 1.032 | 19.71 | +0.13 | d |
| 2 47 20.03 | +0 49 25.1 | 0.584 | 19.01 | +0.08 | d |
| 2 47 39.76 | +1 29 41.1 | 2.054 | 19.77 | −0.02 | d |

| α | δ | z | B' | IDX | Ref. | α | δ | z | B' | IDX | Ref. |
|---|---|---|---|---|---|---|---|---|---|---|---|
| 2 47 45.21 | +0 18 39.0 | 2.015 | 20.08 | +0.01 | d | 2 53 39.85 | +0 27 37.1 | 0.916 | 19.13 | +0.02 | d |
| 2 47 57.22 | −0 20 23.1 | 1.458 | 18.16 | +0.02 | d | 2 53 44.04 | −1 38 42.2 | 0.878 | 16.87 | +0.03 | a |
| 2 48 03.30 | +1 05 49.8 | 1.828 | 19.83 | +0.11 | d | 2 53 45.96 | −0 57 05.1 | 0.720 | 19.01 | +0.62 | e |
| 2 48 05.66 | −0 59 59.9 | 1.845 | 18.64 | +0.07 | d | 2 54 07.94 | −1 37 48.1 | 2.684 | 19.30 | +0.00 | a |
| 2 48 06.79 | −1 00 10.3 | 2.422 | 20.18 | +0.02 | e | 2 54 10.86 | +0 00 43.5 | 2.242 | 18.32 | +0.00 | d |
| 2 48 14.96 | −0 10 12.8 | 0.766 | 19.01 | +0.12 | d | 2 54 24.21 | +0 42 45.8 | 1.115 | 19.67 | +0.04 | d |
| 2 48 23.20 | +0 54 35.0 | 1.708 | 19.45 | +0.05 | d | 2 54 26.14 | +1 26 12.6 | 1.793 | 19.62 | +0.10 | d |
| 2 48 26.70 | +0 35 43.5 | 0.828 | 19.57 | +0.06 | d | 2 54 29.28 | −1 14 18.8 | 0.876 | 19.77 | +0.00 | d |
| 2 48 28.94 | −0 06 18.9 | 1.435 | 18.93 | +0.11 | d | 2 54 32.36 | −0 22 54.8 | 1.585 | 19.65 | −0.04 | d |
| 2 48 34.91 | +1 30 39.3 | 0.815 | 20.00 | +0.06 | d | 2 54 40.26 | −1 13 58.7 | 1.866 | 19.43 | +0.01 | d |
| 2 48 55.43 | −0 39 09.1 | 2.329 | 19.72 | +0.06 | d | 2 54 40.98 | +1 13 39.0 | 1.089 | 19.73 | −0.03 | d |
| 2 49 12.07 | −0 58 57.3 | 1.383 | 19.94 | +0.01 | d | 2 54 43.83 | −0 57 36.9 | 1.032 | 19.92 | +0.05 | d |
| 2 49 13.67 | −0 52 52.9 | 0.817 | 20.11 | +0.04 | d | 2 54 51.41 | −0 10 46.2 | 1.250 | 19.58 | +0.20 | d |
| 2 49 15.52 | −0 58 55.1 | 1.569 | 19.38 | +0.03 | d | 2 54 53.37 | +0 03 45.5 | 1.601 | 19.93 | +0.00 | c |
| 2 49 16.86 | +0 45 22.1 | 1.824 | 20.22 | +0.20 | d | 2 55 13.65 | −1 31 46.7 | 1.520 | 18.45 | +0.09 | d |
| 2 49 17.04 | +1 18 40.2 | 2.981 | 19.54 | −0.02 | e | 2 55 17.59 | +0 08 46.6 | 1.498 | 19.36 | −0.02 | d |
| 2 49 21.88 | +0 44 49.6 | 0.470 | 18.69 | +0.04 | d | 2 55 28.50 | +1 52 05.2 | 1.623 | 19.85 | −0.01 | d |
| 2 49 36.07 | −0 06 27.3 | 2.099 | 19.59 | +0.07 | d | 2 55 30.75 | −0 22 58.4 | 1.557 | 19.82 | +0.10 | d |
| 2 49 42.41 | +2 22 57.2 | 2.805 | 18.93 | −0.01 | a | 2 55 41.94 | −0 15 32.0 | 1.318 | 19.49 | +0.05 | d |
| 2 49 46.50 | +0 15 19.3 | 1.678 | 19.95 | +0.05 | d | 2 55 45.79 | −0 20 04.0 | 2.094 | 19.91 | +0.04 | d |
| 2 49 47.27 | −0 06 16.5 | 0.810 | 17.39 | +0.00 | d | 2 56 11.30 | −1 07 08.9 | 0.905 | 19.43 | +0.15 | d |
| 2 49 54.48 | +0 18 53.6 | 1.106 | 19.09 | +0.08 | d | 2 56 12.61 | +1 50 08.6 | 0.706 | 19.88 | +0.06 | d |
| 2 49 55.30 | +0 48 33.7 | 2.010 | 19.38 | +0.03 | d | 2 56 14.72 | +1 40 28.8 | 0.608 | 18.86 | +0.04 | d |
| 2 50 04.74 | −0 58 42.9 | 1.007 | 19.83 | +0.12 | d | 2 56 20.64 | +0 30 25.9 | 1.569 | 20.43 | −0.04 | d |
| 2 50 04.98 | −1 06 21.5 | 0.846 | 19.95 | +0.19 | d | 2 56 31.80 | −0 00 33.3 | 3.367 | 18.74 | −0.02 | a |
| 2 50 13.01 | +1 40 49.4 | 2.637 | 18.94 | +0.01 | d | 2 56 33.09 | −0 03 57.5 | 2.381 | 19.71 | +0.06 | d |
| 2 50 24.34 | −1 14 34.8 | 1.251 | 19.50 | +0.03 | d | 2 56 37.05 | −0 34 34.7 | 0.361 | 18.23 | +0.02 | d |
| 2 50 34.48 | +1 08 19.5 | 1.331 | 20.16 | +0.18 | d | 2 56 47.40 | +1 46 56.6 | 1.016 | 19.88 | +0.09 | d |
| 2 50 40.67 | +2 03 21.0 | 1.393 | 18.86 | +0.03 | b | 2 56 48.18 | +0 46 35.0 | 1.853 | 19.26 | +0.15 | d |
| 2 50 40.86 | −0 51 15.6 | 0.889 | 19.55 | +0.13 | d | 2 56 55.14 | −0 31 54.0 | 1.998 | 17.71 | +0.01 | d |
| 2 50 41.90 | +0 55 47.0 | 1.030 | 19.60 | +0.01 | d | 2 57 00.45 | −0 37 11.3 | 1.748 | 18.73 | +0.21 | d |
| 2 50 47.21 | −1 46 26.6 | 0.673 | 19.15 | +0.02 | e | 2 57 02.25 | −0 20 56.4 | 1.298 | 19.71 | +0.17 | d |
| 2 50 49.92 | −0 09 42.1 | 1.214 | 19.69 | +0.08 | d | 2 57 03.26 | +0 25 42.7 | 0.535 | 16.82 | +0.01 | d |
| 2 50 51.40 | −0 39 08.3 | 1.363 | 19.87 | +0.08 | d | 2 57 06.51 | −1 09 46.4 | 0.661 | 19.92 | +0.08 | d |
| 2 50 54.00 | −1 46 51.3 | 2.550 | 19.27 | −0.02 | e | 2 57 15.43 | −0 10 13.7 | 1.710 | 19.83 | +0.09 | d |
| 2 50 54.55 | +1 54 32.2 | 1.925 | 19.46 | +0.16 | a | 2 57 23.70 | +0 23 01.6 | 0.820 | 19.69 | +0.05 | d |
| 2 50 58.05 | +0 04 12.6 | 1.810 | 19.40 | +0.02 | d | 2 57 43.13 | +1 16 46.2 | 1.356 | 18.80 | +0.11 | d |
| 2 51 07.14 | −0 01 01.7 | 1.677 | 18.73 | +0.08 | d | 2 57 50.00 | +1 54 23.9 | 1.085 | 19.26 | +0.03 | d |
| 2 51 12.18 | −0 59 17.7 | 2.449 | 18.93 | +0.02 | d | 2 57 53.96 | +2 28 59.6 | 0.115 | 16.23 | +0.17 | c |
| 2 51 22.28 | −0 01 13.5 | 1.688 | 19.84 | +0.08 | d | 2 57 54.15 | −1 00 39.7 | 2.006 | 19.31 | +0.02 | d |
| 2 51 23.31 | −0 23 34.7 | 0.757 | 19.88 | +0.27 | d | 2 57 56.08 | −0 07 30.0 | 0.761 | 19.81 | −0.01 | d |
| 2 51 27.40 | +0 17 05.5 | 1.986 | 19.67 | +0.04 | d | 2 57 59.68 | +0 31 33.7 | 0.806 | 19.86 | +0.11 | d |
| 2 51 49.09 | −0 04 10.6 | 1.213 | 19.96 | +0.10 | d | 2 58 02.48 | −0 27 23.6 | 1.435 | 18.74 | +0.03 | d |
| 2 51 53.15 | −1 53 34.9 | 1.422 | 19.85 | +0.03 | e | 2 58 03.72 | +2 09 26.9 | 1.551 | 19.45 | +0.04 | c |
| 2 51 59.35 | −1 01 42.1 | 1.955 | 20.28 | +0.00 | d | 2 58 07.68 | +0 20 47.2 | 1.112 | 19.21 | −0.01 | d |
| 2 51 59.44 | −0 54 29.1 | 0.433 | 18.01 | +0.12 | d | 2 58 10.43 | +2 10 54.7 | 2.521 | 18.26 | −0.01 | a |
| 2 52 08.14 | +1 41 09.8 | 0.620 | 18.01 | +0.00 | d | 2 58 11.37 | +0 05 06.6 | 1.727 | 19.23 | −0.02 | d |
| 2 52 31.66 | +0 13 15.5 | 0.354 | 17.84 | +0.29 | d | 2 58 11.53 | +0 09 42.8 | 1.497 | 19.88 | +0.12 | d |
| 2 52 39.29 | −0 05 27.3 | 1.885 | 19.68 | +0.50 | d | 2 58 14.54 | +0 42 50.7 | 0.661 | 19.00 | +0.18 | d |
| 2 52 40.10 | +1 36 21.8 | 2.457 | 18.18 | +0.03 | d | 2 58 14.72 | +1 37 06.2 | 1.302 | 19.65 | +0.01 | d |
| 2 52 55.32 | −0 14 25.5 | 1.426 | 19.82 | +0.01 | d | 2 58 25.73 | +1 37 39.1 | 0.595 | 19.03 | +0.13 | d |
| 2 53 12.89 | +0 26 09.5 | 0.921 | 19.11 | +0.11 | d | 2 58 54.46 | +1 45 50.4 | 1.349 | 19.93 | −0.03 | d |
| 2 53 25.53 | +0 41 06.9 | 0.847 | 18.82 | +0.01 | d | 2 59 02.77 | +1 12 53.7 | 2.316 | 19.34 | −0.03 | d |
| 2 53 27.93 | −1 27 48.2 | 1.260 | 19.20 | +0.25 | e | 2 59 03.14 | +1 26 27.3 | 1.578 | 19.20 | +0.04 | d |
| 2 53 28.19 | +0 40 50.8 | 0.531 | 19.40 | +0.07 | d | 2 59 06.29 | +1 04 03.8 | 1.770 | 19.37 | −0.02 | d |
| 2 53 32.65 | +0 58 34.3 | 1.347 | 19.22 | +0.08 | d | 2 59 27.64 | +1 34 31.6 | 1.745 | 18.75 | +0.00 | d |
| 2 53 34.59 | +1 44 30.0 | 1.439 | 19.39 | +0.12 | d | 2 59 33.14 | −0 13 07.0 | 0.641 | 19.46 | +0.48 | d |
| 2 53 39.27 | +0 03 04.2 | 2.012 | 20.27 | +0.09 | e | 2 59 41.16 | −0 10 20.1 | 1.179 | 19.96 | +0.12 | c |

| α | δ | z | B' | IDX | Ref. |
|---|---|---|---|---|---|
| 2 59 46.98 | −0 34 08.0 | 0.706 | 18.72 | +0.01 | c |
| 2 59 58.19 | −1 39 31.9 | 1.520 | 18.37 | +0.07 | c |
| 3 00 39.60 | −0 26 39.9 | 0.693 | 19.17 | +0.46 | c |
| 3 00 42.13 | −0 18 44.5 | 0.707 | 18.29 | +0.00 | c |
| 3 01 07.73 | −0 35 02.9 | 3.205 | 18.67 | −0.01 | c |
| 3 01 08.79 | +0 15 19.5 | 1.656 | 18.61 | +0.02 | c |

References

a- Cristiani et al. 1991
b- Barbieri & Cristiani 1986
c- Véron-Cetty & Véron 1993
d- La Franca et al. 1992
e- This work

**Table 2.** List of the Plates

| Plate | Exp (min) | Date | Epoch/date | Limit (mag) |
|---|---|---|---|---|
| B7221 | 60 | 1981 Sep 28 | 1 (1981.74) | 20.3 |
| B5411 | 60 | 1983 Dec 07 | 2 (1983.94) | 21.1 |
| B5415 | 60 | 1983 Dec 08 | 2 (1983.94) | 20.9 |
| B6321 | 60 | 1986 Jan 05 | 3 (1986.02) | 20.9 |
| B6325 | 60 | 1986 Jan 06 | 3 (1986.02) | 20.9 |
| B6673 | 60 | 1986 Oct 25 | 4 (1986.82) | 21.1 |
| B6692 | 60 | 1986 Nov 21 | 5 (1986.89) | 21.1 |
| B6983 | 60 | 1987 Aug 20 | 6 (1987.65) | 20.9 |
| B6984 | 60 | 1987 Aug 20 | 6 (1987.65) | 20.9 |
| B6989 | 60 | 1987 Aug 24 | 6 (1987.65) | 20.9 |
| B6990 | 60 | 1987 Aug 24 | 6 (1987.65) | 20.9 |
| B6993 | 60 | 1987 Aug 25 | 6 (1987.65) | 20.5 |
| B6994 | 60 | 1987 Aug 25 | 6 (1987.65) | 20.5 |
| B7177 | 60 | 1987 Dec 20 | 7 (1987.97) | 20.9 |
| B7703 | 60 | 1988 Nov 28 | 8 (1988.91) | 20.9 |
| B8268 | 75 | 1989 Oct 24 | 9 (1989.82) | 20.7 |
| B8280 | 75 | 1989 Oct 27 | 9 (1989.82) | 20.9 |
| B8742 | 70 | 1990 Sep 14 | 10 (1990.71) | 20.7 |
| B8746 | 70 | 1990 Sep 21 | 10 (1990.71) | 19.9 |
| B9405 | 75 | 1991 Aug 13 | 11 (1991.62) | 20.7 |
| B9523 | 75 | 1991 Oct 31 | 12 (1991.83) | 20.7 |

to the sky. The observing time for each field was 1 hour, split into two half-hour exposures performing object-sky flipping. A S/N ratio per resolution element larger than 10 was obtained, allowing clear identification for more than 90 per cent of the QSO candidates.

### 2.2. Calibration of the photographic material and error estimation

21 plates taken with the ESO La Silla and UK Schmidt telescopes have been analysed. The combination IIa-O + GG385 filter was always used, defining a passband close to the Johnson $B$. The color transformation between these "natural" $B'$ photographic magnitudes and the Johnson $B$ is (Blair & Gilmore 1982)

$$B' = B - 0.11(B - V). \qquad (1)$$

The field of the sky investigated is included into the limits of the Selected Area 94 (1950.0 coordinates in the following intervals: $2^h43^m27.5^s < \alpha < 3^h1^m34.0^s$ and $-2°05'8.9'' < \delta < 2°46'51.2''$), covering an area of 22.03 square degrees. In Table 2 a detailed list of plates is given (the meaning of *limit-mag* is defined below).

The plate material has been scanned with the COSMOS microdensitometer (MacGillivray & Stobie 1984).

The resulting tables, one per each plate, containing the instrumental magnitudes and other useful parameters for the objects detected, have been merged together in one table. Only objects with at least 4 detections in the first 10 plates have been accepted in this final table. The astrometric error box individuating a common detection has been defined as a circle of 3.5 arcsec of radius. In this way spurious detections (plate defects) are minimized to an acceptable level, while real measurements are in practice never discarded.

To put all the plates in a common magnitude scale:

1) the instrumental magnitudes of each plate separately have been calibrated using polynomial regressions obtained via 116 photometric standards available in the literature or observed specifically for this programme, covering the interval $6.5 < B < 21.2$.

2) for each object for which at least 5 measurements were available, the median of the calibrated magnitudes has been computed, defining a set of *reference magnitudes* ($B'_{ref}$, 48 863 objects).

3) to obtain a useful parameter for the separation of extended from point-like sources, we have computed for each plate and each object the difference between the measured FWHM and the mode of the FWHM distribution at the magnitude of the object. This quantity has then been averaged for each object on the first ten plates.

4) for each plate, the magnitudes computed in step 1) have been re-calibrated against the reference magnitudes, using only point-like objects (41 151 objects).

5) a procedure of uniformization of the usually spatially variable response of the photographic plates has been applied. Each plate has been subdivided in $10 \times 10$ sub-areas, for each of them the differences between the reference and the individual plate magnitudes have been computed and their distribution analysed. The zero-point shifts estimated in this way for each sub-area as a function of the magnitude have been smoothed and applied to the re-calibrated magnitudes. In the following we will refer to the magnitudes obtained in this way as $B'_{final}$.

For each plate the limit of completeness has been assumed to coincide with the maximum of the histogram of last column of Table 2).

For each plate the uncertainties on the $B'_{final}$ magnitudes have been estimated by analysing the distribution of the differences $\Delta B = B'_{final} - B'_{ref}$ as a function of the reference magnitudes for all the point-like objects.

## 3. The SA57 and SGP samples

The SA57 sample is made up of 23 QSOs, studied by Trevese et al. (1994) using 14 plates covering a time-base of 15 years. To evaluate the magnitude errors pertaining to each object we have applied a regression to the data of Table 4 of Trevese et al.'s paper, showing the dependence of the photometric error as a function of the apparent magnitude. For the structure function and variability index calculation (see below) we have considered the same subdivision in 11 epochs applied by Trevese and collaborators.

The South Galactic Pole sample, comprising 283 QSOs, has been studied by Hook et al. (1994) using 11 plates, covering a time interval of 16 years. We have subdivided the 11 plates in 7 epochs, the same as in the original paper (Hook et al. 1994). To evaluate the magnitude errors, the same procedure as in the case of the SA57 sample has been applied (see Table 2 of Hook et al.'s paper).

## 4. Statistical indices of the variability

### 4.1. The structure function

One of the simplest and most immediate methods to analyse the variability of an object is to calculate the structure function (SF). Properties and limitations of the SF and its application to derive timescales and amplitudes of the QSO variability have been described by Simonetti et al. 1985, Trevese et al. 1994, Vio et al. 1992 and references therein. For a finite sequence of measurements $f(t), t = 1, 2, \ldots, N$:

$$SF(\tau) = \frac{1}{N(\tau)} \sum w(t)w(t+\tau)[m(t+\tau) - m(t)]^2 \quad (2)$$

where $m(t)$ are the magnitudes at the time $t$, $w(t)$ are the weights, equal to 1 if a measure exist for the time $t$, 0 otherwise, and

$$N(\tau) = \sum w(t)w(t+\tau) \quad (3)$$

In the present case it is not possible to study in detail the variability of the individual QSOs, due to the relatively small number of plates. We can instead treat the objects in our sample as representative of the QSO "class" and compute a quantity analogous to the SF:

$$SF_e(\tau) = \langle [mag(t) - mag(t+\tau)]^2 \rangle \quad (4)$$

where the $\tau$ is evaluated in the rest-frame of each QSO and the brackets " $\langle \ldots \rangle$ " indicate a mean over the ensemble.

of this approach.

Equation (4) does not provide an optimal estimation of the $SF$ in the sense of statistical "robustness". It depends on the second moment of the distribution of the magnitude differences, which is exposed to the influence of possible "outliers". In principle an estimator based on the mean absolute deviation is expected to give better results, but the presence of the measurement errors (whose subtraction is a necessary condition to obtain a "sample-independent" $SF$) makes its computation complex. To this end, we adopted the procedure described in the appendix.

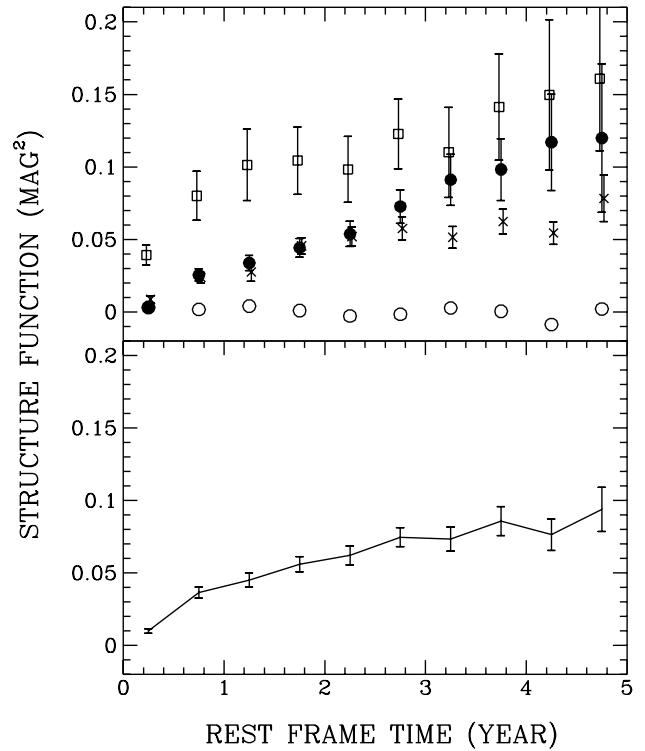

**Fig. 1.** Upper panel: structure functions for the QSO samples SA94 (filled circles), SGP (crosses) and SA57 (open squares). The open circles show the structure function evaluated on the control sample (see text). Lower panel: global structure function obtained from the three QSO samples

Different scenarios require different types of computation of the $SF$. If the variability is assumed to be intrinsic, as in the black-hole or the starburst model, it has to be evaluated in the QSOs restframe: time intervals between pairs of epochs have to be scaled by the $(1 + z_{qso})^{-1}$ cosmological factor. On the other hand, Hawkins (1993) has recently suggested an "intervening" origin for the QSO variability, with nearly all QSOs being microlensed. Ac-cording to him the redshift distribution of the lensing objects (for a uniform population of lenses) is expected to strongly peak at $z \simeq 0.5$, with only weak dependence on the redshift of the source, and therefore the observer's rest-frame is the natural choice for the $SF$ computation.

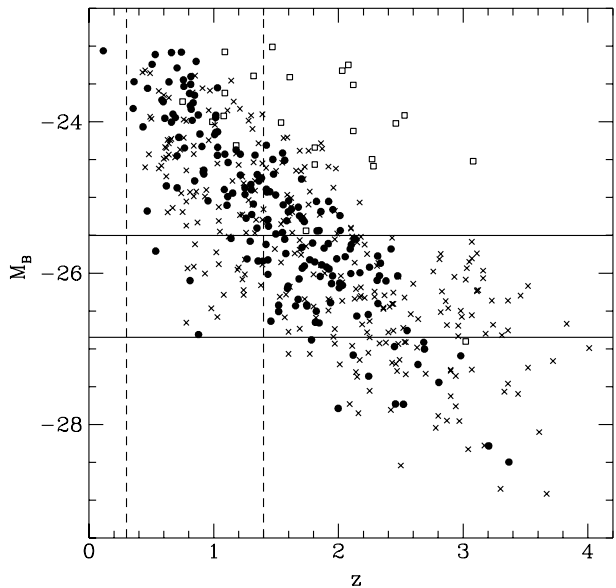

**Fig. 2.** QSO distribution in the $M_B - z$ plane: SA94 (filled circles), SGP (crosses), SA57 (open squares). Absolute magnitudes have been calculated according to Cristiani & Vio (1990)

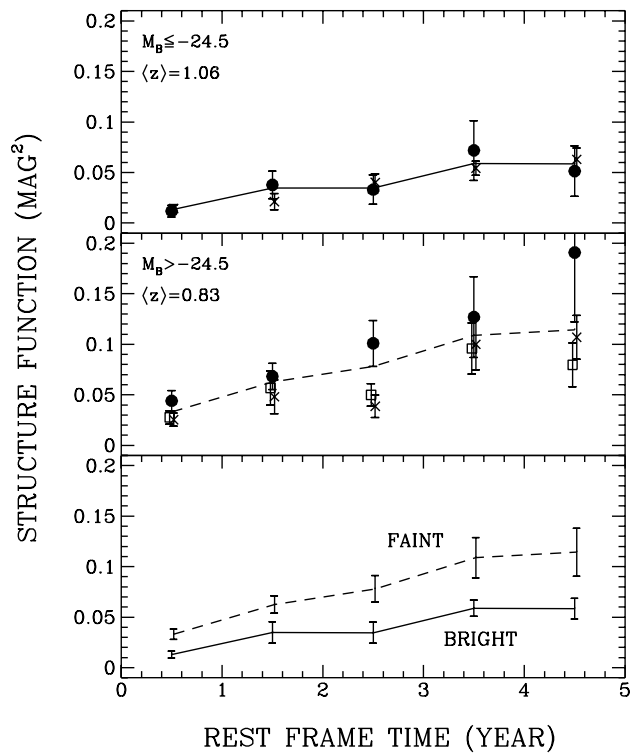

**Fig. 3.** The structure function for QSOs of different luminosity in the redshift interval $0.3 < z < 1.4$. Upper panel: QSOs with $M_B \leq -24.5$. Middle panel: QSOs with $-24.5 < M_B$. Lower panel: comparison between the two SFs. Symbols are the same as in Fig.1

4.1.1. The structure function in the QSOs rest frame

In the lower panel of Fig. 1 the rest frame $SF$ computed *simultaneously* over the SA94, SGP and SA57 samples, with their error-bars, is shown. We observe a steady rise, with a tendency towards flattening, reaching values corresponding to about 0.1 mag$^2$. We stress here that the shape of the ensemble structure function is the result of the combination of the intrinsic variability properties of the QSOs that contribute in different proportions to each bin of time (for example the larger $\Delta t$'s are populated mainly by lower-redshift objects) and is truly representative only if all the QSOs in the samples behave more or less in the same way. This is not the case, as will be shown below.

In the upper panel of Fig. 1 the $SFs$ for the SA94, the SA57 and the SGP samples are shown. The structure function for a control sample in SA94, made of known stars and point-like objects lacking a spectroscopic classification (41 151 objects), is reported too. While the $SFs$ for the SA94 and SGP samples are very similar for $t < 3$yr, they diverge for longer periods of times. Besides, the QSOs in SA57 present a remarkably larger $SF$ for timescales $t <$ 3yr. As indicated by previous works (Cristiani et al. 1990, Hook et al. 1994, Trevese et al. 1994), differences in the $SFs$ may be due to correlations between the variability and physical parameters like absolute luminosity and/or redshift, combined with a different coverage of the $L - z$ plane in the various samples. In Fig. 2, the distribution of all QSOs in the redshift-magnitude plane is illustrated. We see that SA57 objects are, on average, less luminous than the others.

To further investigate and disentangle the above mentioned dependences we have examined two subsamples. The first one is defined within the redshift limits $0.3 < z < 1.4$ (the area between the dashed lines in Fig. 2). The $SF_e$ of less luminous QSOs ($M_B > -24.5$) shows a larger amplitude (Fig. 3) with respect to the more luminous ones ($M_B \leq -24.5$). The second subsample is defined within the absolute luminosity limits $-26.85 < M_B < -25.5$ (the area between the continuous lines in Fig. 2). The $SF_e$ for low-redshift QSOs ($z < 2$) is not distinguishable, within the errors, from the one of high-redshift QSOs ($z \geq 2$), as shown in Fig. 4. No obvious change is apparent neither of

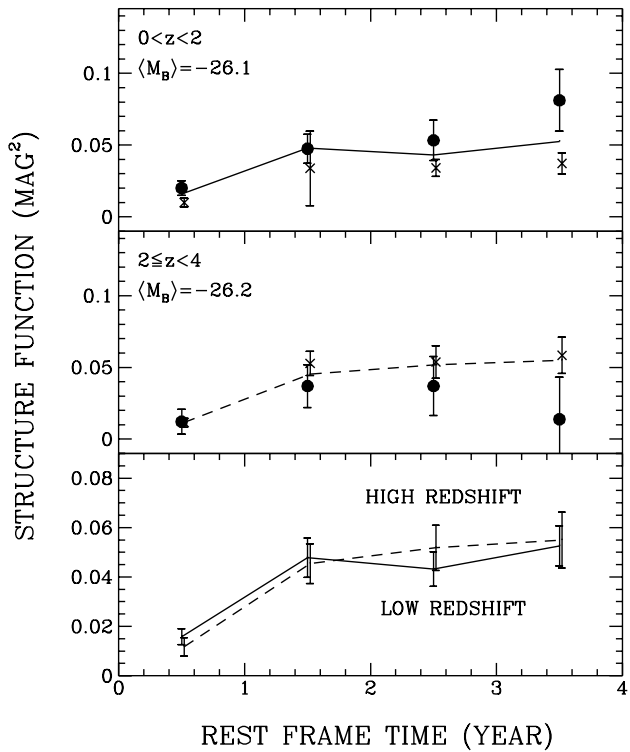

**Fig. 4.** The structure function for QSOs of different redshifts in the luminosity interval $-26.85 < M_B < -25.5$. Upper panel: QSOs with $z < 2$. Middle panel: QSOs with $2 \leq z$. Lower panel comparison between the two SFs. Symbols are the same as in Fig.1

the amplitude nor of the timescales of the variability as a function of the redshift.

To further investigate the dependences of the $SF$ on the absolute magnitude and redshift we have subdivided the $L - z$ plane in a number of sub-areas and evaluated the $SF_e$ in each of them. The result is shown in Fig. 5.

Going from upper to lower panels within the various redshift slices, the anti-correlation between the $SF$ amplitude and absolute magnitude is apparent. A much weaker dependence (if any) exists on the redshift.

In order to parameterize the observed properties of QSO variability with luminosity, redshift and rest-frame time interval, we considered some simple functional forms in which the dependences are separable. Two basic forms of the $SF$ have been explored, a negative exponential $SF = A(1 - e^{-t/\tau})$ and a power-law $SF = At^\tau$. The amplitude $A$ and the timescale $\tau$ have been in turn assumed to be a linear function of the absolute magnitude or redshift, producing six different models.

$$SF = [a + b(M_B + 25.7) + cz](1 - e^{-t/\tau}) \quad (5)$$

Model B:

$$SF = [a + b(M_B + 25.7) + cz]t^\tau \quad (6)$$

Model C:

$$SF = [a + b(M_B + 25.7)](1 - e^{-t/[\tau + c(1+z)]}) \quad (7)$$

Model D:

$$SF = [a + b(M_B + 25.7)]t^{[\tau + c(1+z)]} \quad (8)$$

Model E:

$$SF = (a + cz)(1 - e^{-t/[\tau + b(M_B + 25.7)]}) \quad (9)$$

Model F:

$$SF = (a + cz)t^{[\tau + b(M_B + 25.7)]} \quad (10)$$

Each model was used to fit the binned data shown in Fig. 5 with a non-linear least-squares method. The bin at $< M_B > = -24.3$ and $< z > = 2.28$, based on only 4 objects pertaining to the SA57, appears highly discrepant from the general trend and has been excluded from the fitting procedure (see the next subsection for further comments on this point). The results are given in Table 3.

Errors quoted for the parameters are 68 per cent confidence intervals. Model A and E give the best fit to the data. In both cases dropping the dependence on the redshift provides a significantly worse result. The F-test on the inclusion of the parameter $c$ gives respectively a $5 \cdot 10^{-3}$ and a $2 \cdot 10^{-2}$ probability that the improvement obtained including the z-dependence is due to chance.

### 4.1.2. The structure function in the observer's rest frame

The same parameterizations of the $SF$ have been investigated in the observer's frame, to carry out a comparison with the microlensing model of the QSO variability (Hawkins 1993, Lacey 1994, Alexander 1995). The results are given in Table 4.

Now model A and C give the best fit to the data. In both cases the dependence on the redshift is very weak and dropping it does not change the goodness of fit (model A and C coincide when the dependence on $z$ is dropped). Thus, when analysing the variability in the observer's frame, the best fit is obtained with only three parameters. The amplitude of the $SF$ depends on the absolute magnitude, while the timescale seems to be independent on absolute magnitude or redshift.

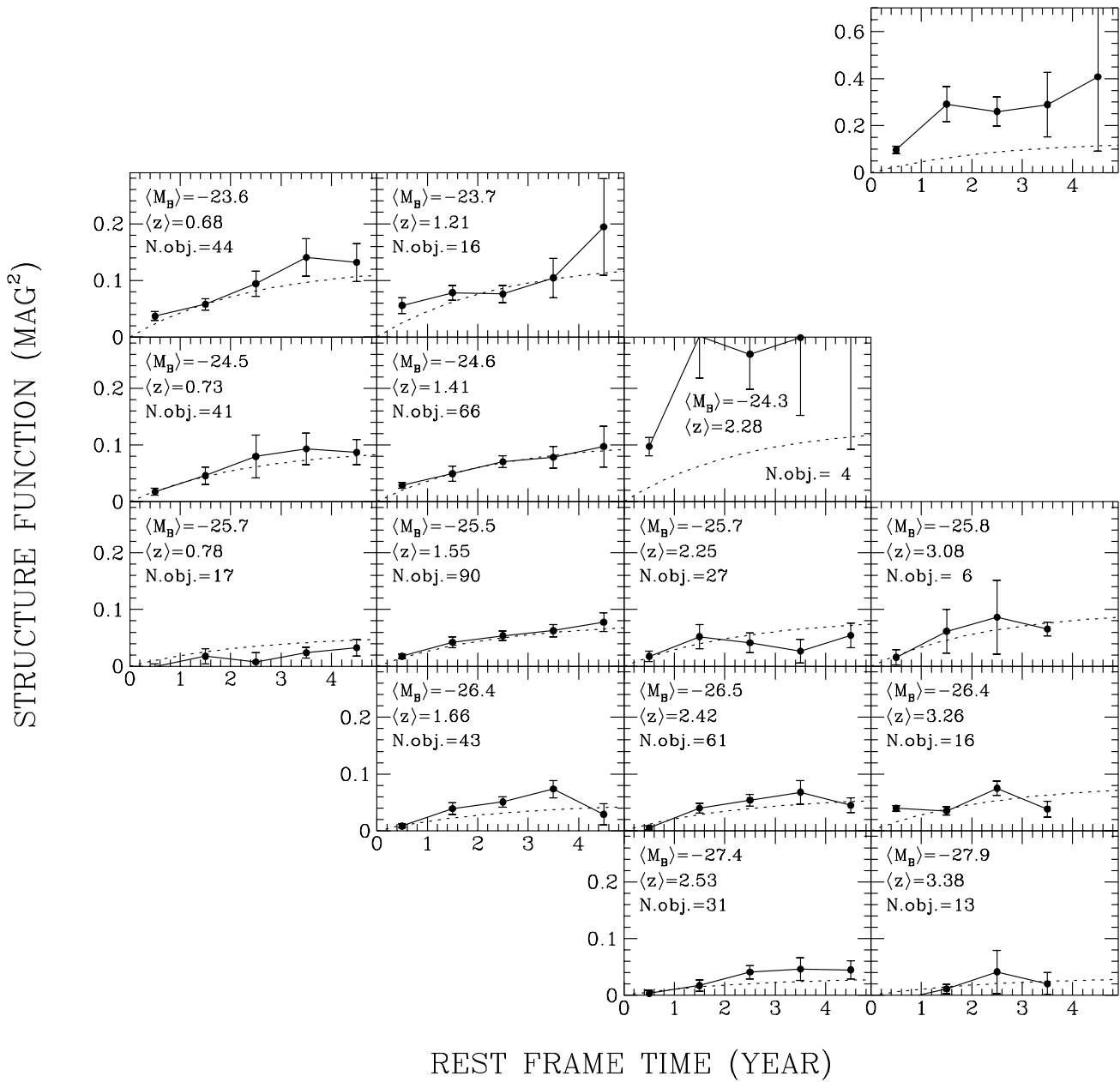

**Fig. 5.** The SF (continuous line) as a function of absolute magnitude and redshift. In the upper left corner of each panel the average absolute magnitude, the average redshift and the number of objects in the bin are shown. The dashed line corresponds to the best-fit model A of Table 3 (see text). The bin at $<M_B> = -24.3$ and $<z> = 2.28$ is re-plotted in the upper right corner with a lower scale in order to show all the points

**Table 3.** Best-fit parameters for the functional forms A, B, C, D, E and F described in the text, evaluated in the QSO rest frame

| Model A ( $\chi^2/\nu = 78/58$ ) | $\tau$ | 2.40 | $\pm 0.09$ |
|---|---|---|---|
| | $a$ | 0.038 | $\pm 0.002$ |
| | $b$ | 0.035 | $\pm 0.013$ |
| | $c$ | 0.021 | $\pm 0.001$ |
| ( $\chi^2/\nu = 92/59$ ) | $\tau$ | 2.34 | $\pm 0.41$ |
| | $a$ | 0.074 | $\pm 0.007$ |
| | $b$ | 0.022 | $\pm 0.023$ |
| | $c$ | 0.000 | *fixed* |
| Model B ( $\chi^2/\nu = 82/58$ ) | $\tau$ | 0.69 | $\pm 0.03$ |
| | $a$ | 0.011 | $\pm 0.001$ |
| | $b$ | 0.012 | $\pm 0.001$ |
| | $c$ | 0.007 | $\pm 0.001$ |
| Model C ( $\chi^2/\nu = 85/58$ ) | $\tau$ | 4.16 | $\pm 0.15$ |
| | $a$ | 0.076 | $\pm 0.003$ |
| | $b$ | 0.028 | $\pm 0.002$ |
| | $c$ | $-0.66$ | $\pm 0.05$ |
| Model D ( $\chi^2/\nu = 96/58$ ) | $\tau$ | 0.44 | $\pm 0.04$ |
| | $a$ | 0.024 | $\pm 0.001$ |
| | $b$ | 0.008 | $\pm 0.002$ |
| | $c$ | 0.106 | $\pm 0.023$ |
| Model E ( $\chi^2/\nu = 79/58$ ) | $\tau$ | 7.1 | $\pm 1.0$ |
| | $a$ | 0.09 | $\pm 0.04$ |
| | $b$ | $-2.75$ | $\pm 0.47$ |
| | $c$ | 0.03 | $\pm 0.02$ |
| ( $\chi^2/\nu = 87/59$ ) | $\tau$ | 4.8 | $\pm 1.0$ |
| | $a$ | 0.11 | $\pm 0.02$ |
| | $b$ | $-1.71$ | $\pm 0.32$ |
| | $c$ | 0.000 | *fixed* |
| Model F ( $\chi^2/\nu = 135/58$ ) | $\tau$ | 0.82 | $\pm 0.04$ |
| | $a$ | 0.02 | $\pm 0.01$ |
| | $b$ | 0.18 | $\pm 0.28$ |
| | $c$ | 0.000 | $\pm 0.002$ |

**Table 4.** Best-fit parameters for the functional forms A, B, C, D, E and F described in the text, evaluated in the observer's frame

| Model A ( $\chi^2/\nu = 81/55$ ) | $\tau$ | 5.4 | $\pm 1.0$ |
|---|---|---|---|
| | $a$ | 0.07 | $\pm 0.01$ |
| | $b$ | 0.026 | $\pm 0.004$ |
| | $c$ | 0.002 | $\pm 0.004$ |
| ( $\chi^2/\nu = 81/56$ ) | $\tau$ | 5.5 | $\pm 0.8$ |
| | $a$ | 0.07 | $\pm 0.01$ |
| | $b$ | 0.026 | $\pm 0.001$ |
| | $c$ | 0.000 | *fixed* |
| Model B ( $\chi^2/\nu = 86/55$ ) | $\tau$ | 0.61 | $\pm 0.05$ |
| | $a$ | 0.015 | $\pm 0.002$ |
| | $b$ | 0.005 | $\pm 0.001$ |
| | $c$ | 0.000 | $\pm 0.001$ |
| Model C ( $\chi^2/\nu = 81/55$ ) | $\tau$ | 5.6 | $\pm 0.8$ |
| | $a$ | 0.073 | $\pm 0.007$ |
| | $b$ | 0.026 | $\pm 0.004$ |
| | $c$ | $-0.06$ | $\pm 0.27$ |
| Model D ( $\chi^2/\nu = 86/55$ ) | $\tau$ | 0.62 | $\pm 0.01$ |
| | $a$ | 0.015 | $\pm 0.001$ |
| | $b$ | 0.005 | $\pm 0.001$ |
| | $c$ | $-0.005$ | $\pm 0.006$ |
| Model E ( $\chi^2/\nu = 85/55$ ) | $\tau$ | 11.0 | $\pm 1.4$ |
| | $a$ | 0.12 | $\pm 0.02$ |
| | $b$ | $-4.06$ | $\pm 0.53$ |
| | $c$ | $-0.01$ | $\pm 0.01$ |
| Model F ( $\chi^2/\nu = 111/58$ ) | $\tau$ | 0.67 | $\pm 0.01$ |
| | $a$ | 0.016 | $\pm 0.001$ |
| | $b$ | 0.13 | $\pm 0.01$ |
| | $c$ | 0.002 | $\pm 0.001$ |

### 4.2. The variability index

Although the union of the 3 samples has increased the number of objects, allowing us to carry out tests on sub-samples, the distribution of the objects in the $L-z$ plane is far from uniform, leaving the possibility for spurious correlations between variability and redshift to be induced by the still surviving correlation between redshift and absolute luminosity. In order to clarify this point we have followed a second approach based on the definition of a variability index for each QSO. By analogy with the $SF$, we have computed a variability index ($IDX$) for each QSO as the intrinsic variance (i.e. taking into account the photometric errors, see appendix) required to reproduce the average absolute differences between pairs of epoch-magnitudes (cfr. eq. A6). The difference with the previous section is that now the intrinsic variance is estimated for each object individually and not at one time collectively for all the objects contributing to a given temporal bin. Only pairs of epochs separated by more than 1 and less than 4 years in the QSOs rest frames have been considered, in order to avoid on the one hand time scales for which the structure function is still quickly rising, as observed in the previous section, and on the other hand time lags unaccessible with the present data for the higher-redshift QSOs. The choice of this time interval is in any case not critical, and the following results depend very weakly on it, as shown by extensive tests. The use of an index defined in this way has an immediate interpretation in terms of a parameterization of the type of Model A (shown in the previous sub-section to be a satisfactory representation of the data), for which the timescale of the rest-frame variability is the same for all the objects and the amplitude varies as a function of absolute magnitude and redshift. Even in the framework of a different parameterization (e.g. Model E) this variability index can still give useful indications, although of less immediate interpretation.

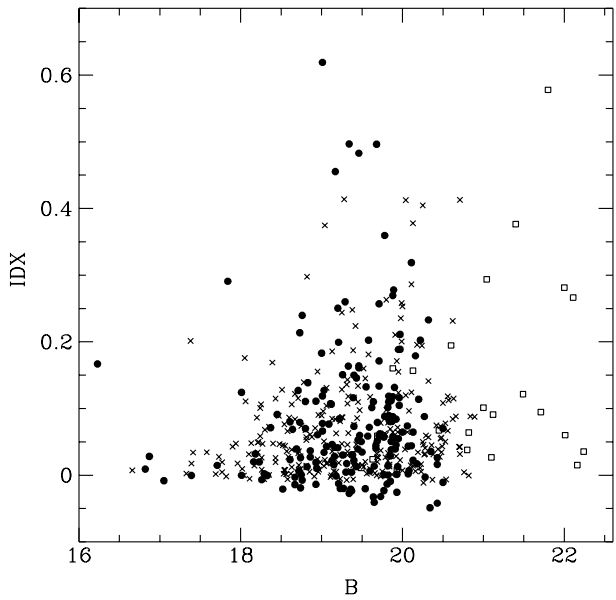

**Fig. 6.** Variability index versus apparent magnitude for the QSOs in the SA94 (•), SGP (×) and SA57 (□)

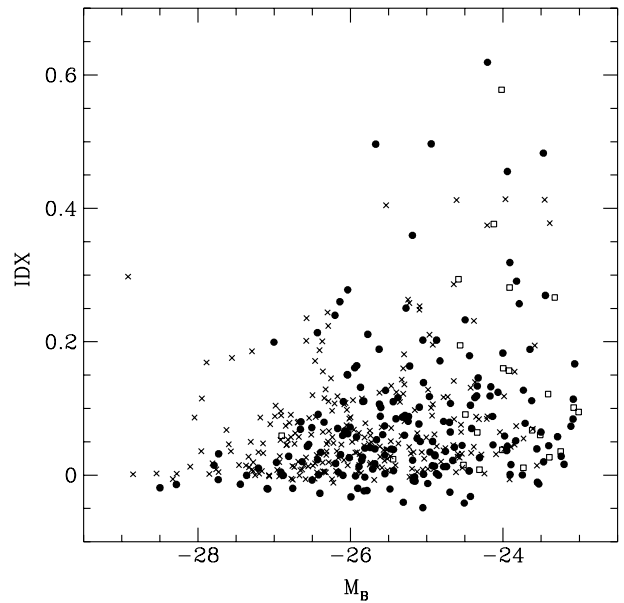

**Fig. 7.** Variability index versus absolute magnitude for the QSOs in the SA94 (•), SGP (×) and SA57 (□)

In Figs. 6, 7, 8 the variability indices vs. apparent magnitude, absolute magnitude and redshift for the QSOs of SA94, SA57 and SGP are shown. The variability indices of the objects in the SA94 are reported in Table 1.

The computation of the variability index for the control sample, defined in the previous sub-section, shows that, as expected, no correlation is present between variability and apparent magnitude, confirming that the measurement errors have been effectively removed.

To quantify the visual impressions given by Figs. 6, 7 and 8 we have analysed the correlation matrix, reported in Table 5.

**Table 5.** Correlation coefficients for SA94 QSOs

|       | $B$            | $z$              | $M_B$          |
|-------|----------------|------------------|----------------|
| $IDX$ | $0.07 \pm 0.07$ | $-0.20 \pm 0.07$ | $0.24 \pm 0.07$ |
| $M_B$ | $0.34 \pm 0.07$ | $-0.83 \pm 0.02$ | —              |
| $z$   | $0.17 \pm 0.07$ | —                | —              |

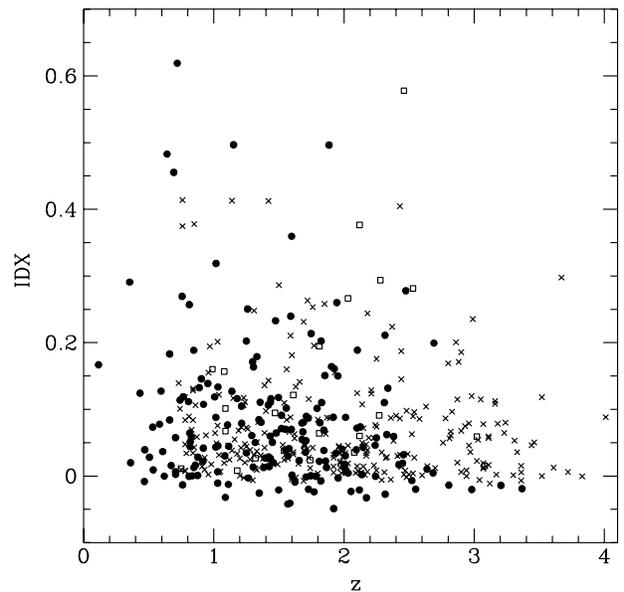

**Fig. 8.** Variability index versus redshift for the QSOs in the SA94 (•), SGP (×) and SA57 (□)

nitude and anticorrelated with redshift (correlation coefficients respectively of +0.24 and −0.20, both with a significance greater than 99 per cent). More luminous and/or higher redshift objects are less variable, but, because of the strong anti-correlation between absolute magnitude and redshift, it is difficult to ascertain if one of the two correlations is spurious. This result is substantially in agreement with previous works (see, for example, Cristiani et al. 1990).

Then, we have merged the SA94 with the SA57 and SGP samples, and re-computed the correlation matrix (Table 6).

**Table 6.** Correlation coefficients for the total of QSOs of the three samples

|     | $B$           | $z$            | $M_B$         |
|-----|---------------|----------------|---------------|
| $IDX$ | $0.21 \pm 0.04$ | $-0.12 \pm 0.04$ | $0.27 \pm 0.04$ |
| $M_B$ | $0.39 \pm 0.04$ | $-0.73 \pm 0.02$ | —             |
| $z$   | $0.29 \pm 0.04$ | —              | —             |

The anticorrelation between absolute magnitude and redshift is reduced. The correlation of the variability index with $M_B$ results unvaried and its significance is increased. The anticorrelation of the variability index with redshift is weaker but marginally significant. It appears now also a correlation with $B$ (0.21 with significance > 99 per cent), a trend caused by the objects of the SA57 sample, less luminous and more variable, as we can see in Fig. 6.

To further investigate the dependences of the variability index on the absolute magnitude and redshift, as in the previous sub-section, we have subdivided the $L - z$ plane in a number of sub-areas and evaluated the average variability index in each of them. The result is shown in Fig. 9.

Going from upper to lower panels within the various redshift slices, the anti-correlation between variability index and absolute magnitude is again apparent. A weaker dependence is observed on the redshift. If we examine the average variability index for the objects with $-25 > M_B > -27$, we find that below redshift 1 $<IDX> = 0.030 \pm 0.009$, while for $z \geq 1$ $<IDX> = 0.069 \pm 0.005$, a difference significant at a $3.8\sigma$ level.

As in the analysis of the $SF$, the bin at $<M_B> = -24.3$ and $<z> = 2.28$, shows a much higher variability than expected from the general trend. Such a discrepant behaviour might be the indication of a bias in the selection technique in favour of variable QSOs. Color selection techniques are based on the separation of the QSO candidates from the stellar locus in a multi-color space (or plane) and a significant bias is expected to occur when (typically for $2.2 < z < 3.5$) a QSO is located adjacent

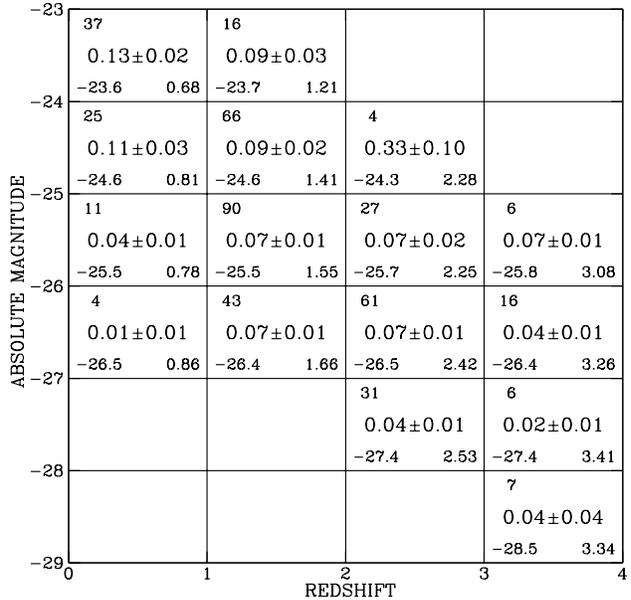

**Fig. 9.** The average rest-frame variability index as a function of absolute magnitude and redshift. In the upper left corner of each panel the number of objects in the bin is shown. In the lower left the average absolute magnitude and in the lower right the average redshift are given. In the middle the average variability index and its $1\sigma$ uncertainty is reported

to, or coincident with, the volume occupied by the stellar locus: QSOs whose detection probabilities are very small, because of variability, may be scattered into regions where they are detectable (e.g. Warren et al. 1994). The bin at $<M_B> = -24.3$ and $2 \leq z < 3$ is based on 4 objects pertaining to the SA57 that have been selected in a $U - J$, $J - F$ plane (Koo & Kron 1988) and are consequently exposed to the above-mentioned bias. A similar bias might also affect the objects in the SGP with $2.2 < z < 3.5$, discovered with multi-colour methods (Warren et al. 1994). We have compared the variability indices of a subsample of the SGP with $2.2 < z < 2.7$ and $-28.0 < M_B < -25.5$ (40 objects) with a corresponding subsample of the SA94 (18 objects) that, being selected with objective prism data, is not affected by this bias. The average variability indices turn out to be $0.069 \pm 0.012$ and $0.066 \pm 0.021$, respectively, showing that for the SGP sample this bias is not playing any significant role.

Another way of disentangling the dependence of the variability on the redshift from the anticorrelation with the absolute luminosity is given by the method of partial correlation analysis (Spiegel 1991). Applying this recipe to the results of Table 6, a value of $0.12 \pm 0.04$ (99 per cent) results for the correlation coefficient between variability index and redshift. Again it appears reasonably well established that, for a given absolute luminosity, higher redshift

(due to the redshift-absolute magnitude anticorrelation) at first glance of the Table 6. The results do not change if one of the samples is eliminated, they simply become less significant, showing that the different time sampling of the different samples does not lead to biases.

The same analysis can be carried out in the observer's rest-frame, estimating the variability indices between 6 and 15 yr. The correlation matrix turns out to be:

**Table 7.** Correlation coefficients computed in the rest-frame of the observer

|     | $B$ | $z$ | $M_B$ |
| --- | --- | --- | --- |
| $IDX$ | $0.10 \pm 0.04$ | $-0.16 \pm 0.04$ | $0.24 \pm 0.04$ |
| $M_B$ | $0.39 \pm 0.04$ | $-0.73 \pm 0.02$ | – |
| $z$ | $0.29 \pm 0.04$ | – | – |

If we remove the dependence of the variability index on the absolute magnitude with the method of partial correlation analysis, the coefficient of correlation between variability index and redshift becomes $+0.02 \pm 0.04$. However, if we examine the average variability index for the objects with $-25 > M_B > -27$, we find that below redshift 1 $<IDX> = 0.043 \pm 0.008$, while for $z \geq 1$ $<IDX> = 0.086 \pm 0.008$, a difference significant at a $3.8\sigma$ level. This latter result seems in contradiction with the analysis of the $SF$ and of the correlation matrix. It has to be considered that it shows a peculiar behaviour of 17 objects at low redshift with respect to 234 objects at high redshift. The $SF$ fitting or the correlation matrix analysis are of course insensitive to such a small subset in the $L - z$ plane (in fact the low-z bins show a high $\chi^2$ per bin).

## 5. Discussion and Conclusions

By merging three different samples (SA94, SGP, SA57), we have been able to analyse the variability of a statistically well defined set of 486 optically selected QSOs.

The ensemble structure function and individual variability indices of the QSOs have been evaluated by means of "robust" statistical estimators, less exposed to the influence of possible "outliers" with respect to more conventional estimators. The effects of the photometric errors have been subtracted, allowing a meaningful comparison of the three different samples.

Although the coverage of the $M_B - z$ plane of the present samples is not completely uniform, the definition of suitable sub-samples has allowed us to disentangle the average relationships between variability and redshift and between variability and luminosity.

structure function and the individual variability indices in the QSOs rest frame show that:

1. A negative correlation between variability and luminosity is clearly present, in the sense that more luminous QSOs show less variability (in magnitude), confirming previous results (Cristiani et al. 1990, Hook et al. 1994).
2. A significant positive correlation exists between variability and redshift, as suggested already by Giallongo et al. (1991).
3. Such correlations may be equally well parameterized either with a model in which the timescale of the variability is fixed for all the QSOs and the amplitude linearly increases with the absolute magnitude and redshift, or with a model in which the timescale of the variability linearly depends on the absolute magnitude and the amplitude is only a function of the redshift.

The same analysis carried out in the observer's frame provides the following results:

1. There is a negative correlation between variability and luminosity.
2. The timescale of variability does not depend significantly neither on the absolute magnitude nor on the redshift.
3. The ensemble structure function is well represented by a parameterization in which, with a fixed timescale of about 5.5 yr, the amplitude linearly increases with the absolute magnitude.
4. Although the general behaviour of the $SF$ does not show a systematic variation of the timescale and/or amplitude with redshift, if we examine the average variability index for objects with $-25 > M_B > -27$, we find that below redshift 1 quasars are significantly less variable than at higher redshift.

The anti-correlation observed in the QSOs rest frame between absolute luminosity and variability is considered by some authors as an evidence in favor of the *sub-units* model, in which variability is caused by an ensemble of individual flares. To test this hypothesis, following Pica and Smith (1993), we have made a simple calculation. Let a QSO be composed by $N$ random flaring sub-units; in this case the signal is proportional to $N$, whereas the noise (e.g. a supernova event in the starburst model) is proportional to $\sqrt{N}$. Then $S/N \approx N/\sqrt{N} = \sqrt{N} \propto \sqrt{L}$, where $L$ is the luminosity of the QSO. Thus the amplitude of the relative variability should decrease as $1/\sqrt{L}$ and the $SF$ as $1/L$. A difference of 2 magnitudes, corresponding to a ratio

$$\frac{L_2}{L_1} = 10^{-0.4(M_1 - M_2)} = 6.3 \tag{11}$$

should give rise to an equal ratio between the amplitudes of the $SF$

$$\frac{SF(M_1)}{SF(M_2)} = \frac{L_2}{L_1} \simeq 6.3 \tag{12}$$

value < 2, unacceptably lower than the predictions of the *sub-units* model, at least in its simplest form. However, any additional background luminosity contributing to the total luminosity of the QSOs would introduce departures from the $L^{-1}$ relationship. This is, indeed, the case of the starburst model for AGNs (Terlevich et al. 1992), in which variability is caused by the supernova explosions of a nuclear young stellar cluster. The stellar background in this model accounts for about half the luminosity in B band (Aretxaga & Terlevich 1994), and tends to flatten the $L^{-1}$ law (Aretxaga 1993; Aretxaga et al. 1993). Even a more standard model of a black-hole which induces pulses of light (due to, for example, accretion events or stellar impacts onto the disk) would introduce some background luminosity in a quiescent stage that, potentially, could modify that law. Another important effect is expected to be due to the variability-redshift correlation. If such a correlation is not accounted for, since the more luminous QSOs are those of higher redshifts, the amplitude of their observed B-band variability would tend to be larger than what expected on the basis of the luminosity-variability anticorrelation, causing a flattening of the a priori $L^{-1}$ law. This effect is important enough to introduce serious departures from the standard law for the sub-unit model, and account for the flattening observed (Cid Fernandes 1995).

Comparisons with more refined models of variability produced by dense stellar clusters and supernova explosions are being developed, and will be discussed in detail in a separate paper. Unfortunately these models are the only ones that, so far, permit a detailed comparison between observed light curves and the physical parameters of the events that produce the variability (energy, metallicity, etc).

The positive variability-redshift correlation may be interpreted as an increase of variability at the shorter and shorter wavelengths redshifted in the fixed observational bandpass (Giallongo et al. 1991), as already observed in individual objects (Edelson et al. 1990, Kinney et al. 1991, Paltani & Courvoisier 1994). Such a behaviour should be connected to the physical mechanisms governing the energy flux at different wavelengths. In order to confirm this result and its interpretation it will be important to study the ensemble variability of the same QSO samples on red plates.

Hawkins (1993) has recently proposed that nearly all QSOs are being microlensed. According to him, since the redshift distribution of the lensing objects (for a uniform population of lenses) is expected to strongly peak at $z \simeq 0.5$, with only weak dependence on the redshift of the source, for such a model one would not expect to see a significant increase in time scales with redshift (in the rest-frame of the observer). The present data are not able to disprove such a behaviour. It remains to be established whether the absence of a dependence of the variability the structure function, is a true result of microlensing or the time dilation effect counterbalances the evolution of the intrinsic variability with redshift. On the other hand, it has also to be clarified if the significant decrease of the variability index of QSOs with $-25 > M_B > -27$ found at $z < 1$ is due to an intrinsic change of the QSO variability or to a smaller expected frequency of the microlensing. In this respect, it will be necessary to carry out further tests, for example on the achromaticity of the variability of individual objects, for which again the study of the ensemble variability of the same QSO samples on red plates will be extremely important.

*Acknowledgements.* It is a pleasure to thank E. Giallongo, R. Terlevich, D. Trevese and F. Vagnetti for enlightening discussions, and I. Hook for providing the data of the SGP before publication. This research has been partially supported by the ASI contracts 92-RS-102 and 94-RS-107 and by ANTARES, an astrophysics network funded by the HCM programme of the European Community. IA acknowledges the Basque Government for the fellowship BFI93.009. RV thanks an ESA research fellowship. SC "thanks" the Italian CNR and the University of Padua for the ludicrous (and decreasing) financial support.

## A. Taking into account measurement errors in the computation of the structure function

The difference between the magnitudes at two epochs, in presence of errors, is given by

$$|m_i + \epsilon_i - m_j - \epsilon_j| \tag{A1}$$

where $m_i$ is the true magnitude of an object at the $i$-th epoch and $\epsilon_i, \epsilon_j$ are the errors. But, because of the triangular inequality,

$$|m_i - m_j| \geq |m_i + \epsilon_i - m_j - \epsilon_j| - |\epsilon_i - \epsilon_j| \tag{A2}$$

where the term at the first member is the difference of the "true magnitudes" we are looking for and cannot be calculated in a straightforward way.

To subtract the influence of the errors we have adopted the following procedure:

1. The various plates have been grouped in *epochs* $t_i$, according to the time of observation (12 for the SA94 sample, see Table 2; 11 for SA57; 7 for SGP).
2. For each object, the median of all the plate-magnitudes has been calculated. To avoid biases, for each plate, the measurements of all the objects with a median magnitude fainter than the plate limit-magnitude have been excluded from the following computations.
3. The *epoch-magnitudes EMAG* have been calculated as the mean of the magnitudes of each epoch.
4. For each QSO the quantity

$$\mu_{ij} = \frac{|EMAG_i - EMAG_j|}{\sqrt{(\sigma_i^2 + \sigma_j^2)}} \tag{A3}$$

has been computed, where $\sigma_i^2$, $\sigma_j^2$ are the magnitude uncertainties at the epochs $i$, $j$. The $\mu_{ij}$ have been assigned to the appropriate temporal bin, identified by

$$\Delta t = \frac{|t_i - t_j|}{1 + z} \tag{A4}$$

when the $SF$ is computed in the QSOs rest-frame or simply by

$$\Delta t = |t_i - t_j| \tag{A5}$$

when the $SF$ is computed in the observer'ss rest-frame.

5. For each temporal bin the average deviation

$$DEV(\Delta t) = \frac{\sum \mu_{ij}}{N} \tag{A6}$$

has been evaluated, where the sum is carried out on the $N$ magnitude differences pertaining to the bin. This quantity $DEV(\Delta t)$ is the result of the magnitude variations due to the intrinsic QSO variability plus the photometric errors of each epoch-magnitude.

6. We have then computed the quantity

$$DEV_s(\Delta t) = \frac{\sum |\nu_{ij}|}{N} \tag{A7}$$

where $|\nu_{ij}|$ is the expected absolute deviation for the epochs $i$ and $j$, assuming that the difference of the epoch-magnitudes of an object is the result of a gaussian process with a variance resulting from the sum of the variances due to the appropriate photometric uncertainties plus a variance $SF_e$ due to intrinsic variability. The correct value of the $SF_e$ in a given temporal bin is assumed to be the one for which

$$DEV - DEV_s = f(SF_e) = 0 \tag{A8}$$

7. Since the differences $|EMAG_i - EMAG_j|$ are not independent measurements, to estimate the errors we have used the bootstrap sampling method (Barrow et al. 1984). The bootstrap method mimics the process of repeating the input samples a large number of times. To generate the artificial samples we selected 486 times a QSO at random from the original samples. Thus the same QSO could appear more than once in the artificial sample. The process was repeated 100 times and for each artificial sample the $SF_e$ was computed. For each time bin of the $SF_e$ the $RMS$ of the estimates obtained from the 100 simulations has been adopted as the uncertainty on the value computed in the Eq. (A8).